\documentclass[12pt,onecolumn,oneside,a4paper,peerreview]{IEEEtran}
\usepackage{cite}
\usepackage{graphicx}
\usepackage{amsmath}
\usepackage{times}
\usepackage{latexsym}
\usepackage{graphicx}
\usepackage{bm}
\usepackage{amssymb}
\usepackage[center]{caption2}
\usepackage{stfloats}
\usepackage{cases}
\usepackage{array}
\usepackage{setspace}
\usepackage{fancyhdr}
\usepackage{citesort}

\renewcommand{\QED}{\QEDopen}
\newtheorem{theorem}{Theorem}
\newtheorem{lemma}{Lemma}
\newtheorem{corollary}{Corollary}

\newtheorem{proposition}{Proposition}
\begin{document}

\title{Performance Analysis of Optimal Single Stream Beamforming in MIMO Dual-Hop AF Systems}
\vspace{0.5cm}
\author{
\begin{minipage}{0.98\columnwidth}
\begin{center}
\authorblockN{Caijun Zhong\authorrefmark{4},
            Tharm Ratnarajah\authorrefmark{4},
            Shi Jin\authorrefmark{3},
            Kai Kit Wong\authorrefmark{2}\\
\vspace*{0.5cm} \small{
\authorblockA{\authorrefmark{4}The Institute of Electronics, Communications and Information Technology, Queen's University Belfast, Belfast}\\
\authorblockA{\authorrefmark{3}National Mobile Communications Research Laboratory, Southeast University, Nanjing, China}\\
\authorblockA{\authorrefmark{2}Department of Electronic and Electrical Engineering, University College London, United Kingdom}
}}
\end{center}
\end{minipage}
} \IEEEaftertitletext{\vspace{-0.75\baselineskip}}

\maketitle \vspace{0.2cm}
\begin{abstract}
This paper investigates the performance of optimal single stream
beamforming schemes in multiple-input multiple-output (MIMO)
dual-hop amplify-and-forward (AF) systems. Assuming channel state
information is not available at the source and relay, the optimal
transmit and receive beamforming vectors are computed at the
destination, and the transmit beamforming vector is sent to the
transmitter via a dedicated feedback link. Then, a set of new
closed-form expressions for the statistical properties of the
maximum eigenvalue of the resultant channel is derived, i.e., the
cumulative density function (cdf), probability density function
(pdf) and general moments, as well as the first order asymptotic
expansion and asymptotic large dimension approximations. These
analytical expressions are then applied to study three important
performance metrics of the system, i.e., outage probability, average
symbol error rate and ergodic capacity. In addition, more detailed
treatments are provided for some important special cases, e.g., when
the number of antennas at one of the nodes is one or large, simple
and insightful expressions for the key parameters such as diversity
order and array gain of the system are derived. With the analytical
results, the joint impact of source, relay and destination antenna
numbers on the system performance is addressed, and the performance
of optimal beamforming schemes and orthogonal space-time
block-coding (OSTBC) schemes are compared. Results reveal that the
number of antennas at the relay has a great impact on how the
numbers of antennas at the source and destination contribute to the
system performance, and optimal beamforming not only achieves the
same maximum diversity order as OSTBC, but also provides significant
power gains over OSTBC.
\end{abstract}

\vspace{0.2cm}

\vspace{0.2cm}
\vskip 2ex \vspace*{0ex}
\begin{tabular}{ll}
{Corresponding Author\;:} {Caijun Zhong}\\
{The Institute of Electronics, Communications and Information Technology}\\
{Queen's University Belfast, Queen's Road, Queen's Island}\\
{Belfast, Northern Ireland, BT3 9DT}\\
{E-mail: caijunzhong@zju.edu.cn} \\
\end{tabular}

\newpage

\section{Introduction}
Multiple-input multiple-output (MIMO) antenna technology can be used to substantially increase the spectral efficiency or significantly improve the reliability of the communication systems \cite{Foschini,Alamouti}, and has become one of the key enabling technologies for future wireless communication systems. Parallel to the development of MIMO technology, dual-hop relaying technology, where an intermediate relay node helps forward the source signal to the intended destination node, is another efficient approach to considerably improve the performance and extend the coverage of the communication systems \cite{M.Hasna1,M.Hasna2}. The huge potential of achieving both the MIMO and relaying benefits has resulted in enormous interest in the MIMO relaying systems \cite{B.Wang,Y.Fan}.

Due to the low complexity of amplify-and-forward (AF) relaying protocol, it is of great importance to understand the performance of MIMO dual-hop AF systems. Prior works on MIMO dual-hop AF systems have largely focused on the asymptotic characterization of the information theoretical measures such as capacity scaling behavior and diversity-multiplexing tradeoff (DMT). For instance, in \cite{H.Bolcskei}, the impact of the number of antennas at the relay on the asymptotic capacity was studied, while \cite{V.Morgen,J.Wagner} investigated the asymptotic capacity of multi-hop AF relay systems using tools from large-dimensional random matrix theory. The DMT of MIMO dual-hop AF systems was established in \cite{S.Yang}, and was later extended to the multi-level AF systems \cite{S.Yang1}.

A few results on MIMO dual-hop AF systems in the finite regime have
also emerged recently. In particular, an exact ergodic capacity
analysis was presented in \cite{S.Jin}, while in
\cite{Y.Song,Pratha}, the symbol error rate (SER) performance of
orthogonal space-time block-coding (OSTBC) schemes in such system
was addressed using a moment generating function based approach. The
performance of optimal beamforming schemes has also been
investigated in various multiple antenna dual-hop AF systems
\cite{R.Louie,D.Costa,J.Kim,H.Min}. However, the major limitation of
all these prior works \cite{R.Louie,D.Costa,J.Kim,H.Min} is that one
or more nodes is restricted to have only one antenna. In addition,
all these works consider the channel state information (CSI)
assisted relaying protocol, which however requires the relay having
access to the CSI. With the implementation of multiple antennas,
acquiring the CSI at the relay entails higher implementation
complexity and cost, hence, it may not
be suitable for certain applications with low cost budgets. On the other hand, the fixed-gain
relaying protocol, where the relay simply forwards a scaled version
of the received signal, becomes more appealing for such
applications. Despite its practical importance, the performance of
optimal beamforming schemes in general MIMO dual-hop AF systems with
fixed-gain relaying protocol remains unknown.

Motivated by this, we adopt the same system model as in
\cite{V.Morgen,Pratha,S.Jin} and study the analytical performance of
optimal beamforming schemes in the general MIMO dual-hop AF systems.
It is worth pointing out that, with multiple antennas at all the
three nodes, it is possible to send multiple independent date
streams at the same time, namely, multi-stream beamforming systems
(See, i.e., \cite{X.Tang,O.Medina} and references therein). In
general, the multi-stream beamforming system provides higher data
rate, but experiences worse error performance compared with the
single stream beamforming system. In the current paper, we focus
exclusively on the single stream beamforming system, and leave the
analysis of the multi-stream beamforming system for future works. In
addition, we assume that the CSI is not available at the source and
relay. Instead, it is estimated at the destination via the methods
proposed in \cite{F.Gao,Y.Rong,T.Kong}. Hence, the optimal receive
and transmit beamforming vectors are computed at the destination,
and the transmit beamforming vector is then forwarded to the
transmitter through a dedicated feedback link.

The main contribution of the paper is the derivation of a set of new
statistical results pertaining to the maximum eigenvalue of the
underlying MIMO dual-hop AF channel matrix. In particular, we
present exact closed-form expressions for the cumulative
distribution function (cdf), probability density function (pdf), and
general moments, as well as asymptotic first order expansion of the
cdf and pdf of the maximum eigenvalue. In addition, the statistical
behavior of the maximum eigenvalue is also characterized in the
asymptotic large antenna regime. We then apply these statistical
results to study the performance of optimal beamfomring systems in
terms of three key performance metrics, i.e., outage probability,
average SER and ergodic capacity. Furthermore, we provide detailed
treatments for certain important special cases, and obtain key
performance indicators such as diversity order, array gain and high
signal-to-noise ratio (SNR) slope and power offset of the system.
Finally, we compare the results of optimal beamforming with OSTBC
schemes, and show that optimal beamforming schemes achieve the same
maximum diversity order as OSTBC schemes, but have significant power
gains over OSTBC schemes.

The remainder of the paper is organized as follows. Section II
introduces the system model. Section III presents a host of new
statistical results of the maximum eigenvalue, while Section IV
gives a detailed investigation on the performance of optimal
beamforming schemes in terms of outage probability, average SER and
ergodic capacity. Finally, Section V concludes the paper.

Throughout the paper, the following notation is adopted. Vectors are represented as columns and are denoted in lower-case bold-face, while matrices are represented in upper-case bold-face. The superscript $(\cdot)^\dag$ indicates the matrix conjugate-transpose operations, respectively. We use $\Gamma(x)$ to denote the gamma function, $\det(\cdot)$ to denote the matrix determinant, $\|\cdot\|$ to denote the Frobenius norm, ${\tt E}\{\cdot\}$ to denote the expectation operation, ${\bf I}_r$ to denote an $r\times r$ identity matrix, and $\mathcal{CN}^{a\times b}$ to denote an $a\times b$ matrix with entries being identically and independently distributed (i.i.d.) zero mean circular symmetric complex Gaussian (ZMCSCG) random variables with unit variance.

\section{System Model}
We employ the same AF MIMO dual-hop system model as in \cite{V.Morgen,Pratha,S.Jin}. Suppose that there are $n_s$ antennas at the source terminal, $n_r$ antennas at the relay and $n_d$ antennas at the destination terminal, which we represent this by the $3$-tuple $(n_s,n_r,n_d)$. All terminals operate in half-duplex mode, and as such communication occurs from source to relay and from relay to destination in two separate time slots. The end-to-end input-output relation of this channel is then given by
\begin{align}\label{eq:signalmodel}
{\bf{y}} = {\bf{H}}_2 {\bf{FH}}_1 {\bf{s}} + {\bf{H}}_2{\bf{Fn}}_{n_r} + {\bf{n}}_{n_d},
\end{align}
where ${\bf s}$ is the transmit symbol vector, ${\bf n}_{n_r}$ and
${\bf n}_{n_d}$ are the noise vectors at the relay and destination
terminals, respectively, ${\bf F}
=\sqrt{\frac{\alpha}{n_r(1+\rho)}}\mathbf{I}_{n_r}$ ($\alpha$
corresponds to the overall power gain of the relay terminal) is the
forwarding matrix at the relay terminal which simply forwards scaled
versions of its received signals\footnote{It is worth pointing out
that, with the implementation of multiple antennas, the variance of
the received power per antenna at the relay is reduced.}, and $
{\bf{H}}_1 \in \mathcal{CN}^{n_r \times n_s}$ and ${\bf{H}}_2 \in
\mathcal{CN}^{n_d \times n_r}$ are the MIMO channel matrices from
the source to relay and from the relay to destination, respectively.
The additive noise at the relay and destination are assumed to be
white in both space and time and are modeled as ZMCSCG with unit
variance, i.e., $ {\tt E}\left\{ {{\bf{n}}_{n_r} {\bf{n}}_{n_r}^\dag
} \right\} ={\bf{I}}_{n_r}$ and $ {\tt E}\left\{ {{\bf{n}}_{n_d}
{\bf{n}}_{n_d}^\dag }\right\} = {\bf{I}}_{n_d}$.

For the transmit beamforming scheme, we have ${\bf s} = \mathbf{w}_tx$ where $\mathbf{w}_t$ is the beamforming vector with unit norm, i.e., $\|\mathbf{w}_t\|^2 = 1$, and $x$ is the transmit symbol satisfying ${\tt E}\{|x|^2\} = \rho$. At the destination node, the receiver combines the signal by multiplying the received signal vector with $\mathbf{w}_r$. Hence, the post-processed signal at the receiver node can be expressed as
\begin{equation}
\mathbf{w}_r^{\dag}{\bf{y}} = \sqrt{a}\mathbf{w}_r^{\dag}{\bf{H}}_2 {\bf{H}}_1 {\bf w}_tx + \sqrt{a}\mathbf{w}_r^{\dag}{\bf{H}}_2{\bf{n}}_{n_r} + \mathbf{w}_r^{\dag}{\bf{n}}_{n_d},
\end{equation}
where $a = \frac{\alpha}{n_r(1+\rho)}$. As a result, the end-to-end SNR is given by
\begin{equation}\label{eqn:snr1}
\gamma =\frac{\mathbf{w}_r^{\dag}\mathbf{H}_2\mathbf{H}_1\mathbf{w}_t\mathbf{w}_t^{\dag}\mathbf{H}_1^{\dag}\mathbf{H}_2^{\dag}\mathbf{w}_ra\rho}{\mathbf{w}_r^{\dag}\left(a\mathbf{H}_2\mathbf{H}_2^{\dag}+\mathbf{I}_{n_d}\right)\mathbf{w}_r}.
\end{equation}
It can be easily observed that the optimal combining vector $\mathbf{w}_r$ in maximizing the SNR is
\begin{equation}\label{eqn:comvec}
\mathbf{w}_r=\left(a\mathbf{H}_2\mathbf{H}_2^{\dag}+\mathbf{I}_{n_d}\right)^{-1}\mathbf{H}_2\mathbf{H}_1\mathbf{w}_t.
\end{equation}
Substituting (\ref{eqn:comvec}) into (\ref{eqn:snr1}), the corresponding maximum SNR is given by
\begin{equation}
\gamma =a\rho\mathbf{w}_t^{\dag}\mathbf{H}_1^{\dag}\mathbf{H}_2^{\dag}\left(a\mathbf{H}_2\mathbf{H}_2^{\dag}+\mathbf{I}_{n_d}\right)^{-1}\mathbf{H}_2\mathbf{H}_1\mathbf{w}_t.
\end{equation}
Therefore, the optimal transmit beamforming vector corresponds to the eigenvector associated with the maximum eigenvalue of $\mathbf{H}_1^{\dag}\mathbf{H}_2^{\dag}\left(a\mathbf{H}_2\mathbf{H}_2^{\dag}+\mathbf{I}_{n_d}\right)^{-1}\mathbf{H}_2\mathbf{H}_1$, giving the resultant SNR as
\begin{equation}
\gamma = a\rho\lambda_{\sf max},
\end{equation}
where $\lambda_{\sf max}$ is the maximum eigenvalue of the random matrix $\mathbf{H}_1^{\dag}\mathbf{H}_2^{\dag}\left(a\mathbf{H}_2\mathbf{H}_2^{\dag}+\mathbf{I}_{n_d}\right)^{-1}\mathbf{H}_2\mathbf{H}_1$.


\section{Statistical Properties of $\lambda_{\sf max}$}
This section presents a set of new statistical results of $\lambda_{\sf max}$, including exact expressions for the cdf, pdf, and general moments of the maximum eigenvalue $\lambda_{\sf max}$, as well as the asymptotic first order expansions. In addition, the statistical behaviors of $\lambda_{\sf max}$ are also investigated in the asymptotic large antenna regimes. For notation convenience, we find it useful to define $q\triangleq \min(n_d, n_r)$, $p\triangleq\max(n_d, n_r)$, $s\triangleq\min(n_s,q)$, $t\triangleq\max(n_s,q)$, $m\triangleq\min(n_s,p)$, $n\triangleq\max(n_s,p)$, $\theta(i,j) \triangleq 2q+p-i-j-s$, and $\tau(i,j)\triangleq s+i+j-q-2$.

\subsection{Exact Expressions for the cdf}
The following theorem gives the cdf of $\lambda_{\sf max}$ and will be useful in the analysis of the outage and SER performance of optimal beamforming scheme in MIMO dual-hop AF systems.
\begin{theorem}\label{theorem:1}
Let $ {\bf{H}}_1  \in \mathcal{C}^{n_r  \times n_s}$ and $ {\bf{H}}_2  \in \mathcal{C}^{n_d  \times n_r}$, and $a$ being a non-negative constant, the cdf of the maximum eigenvalue $\lambda_{\sf max}$ of $\mathbf{H}_1^{\dag}\mathbf{H}_2^{\dag}\left(a\mathbf{H}_2\mathbf{H}_2^{\dag}+\mathbf{I}_{n_d}\right)^{-1}\mathbf{H}_2\mathbf{H}_1$ is given by
\begin{align}\label{eqn:exactcdf}
{\cal F}_{\lambda_{\sf max}}(x) =\frac{(-1)^{n_s(t-n_s)}\det\left(\mathbf{\Phi}(x)\right)}{\prod_{i=1}^{q}\Gamma(q-i+1)\Gamma(p-i+1)},
\end{align}
where $\mathbf{\Phi}(x)$ is a $q\times q$ matrix with entries
\begin{equation}\label{eqn:phi}
[\mathbf{\Phi}(x)]_{i,j}=\left\{\begin{array}{cc}
(-1)^{q-s-i}\sum_{l=0}^{q-i+j-1}{q-i+j-1\choose l}a^l\Gamma(p+l+i-j) & \mbox{for }i\leq q-s,\\
\sum_{l=0}^{\tau(i,j)}{\tau(i,j)\choose l}a^l\Gamma(\theta(i,j)+l+1) - e^{-ax}B(x) & \mbox{for }i> q-s.
\end{array}\right.
\end{equation}
where
\begin{align}
B(x) = 2\sum_{k=0}^{t-i}\frac{x^k}{k!}\sum_{l=0}^{\tau(i,j)+k}{\tau(i,j)+k\choose
l}a^lx^{\frac{\theta(i,j)+l-k+1}{2}}K_{\theta(i,j)+l-k+1}(2\sqrt{x}),
\end{align}
where $K_v(x)$ is the modified Bessel function of the second kind \cite{Table}.
\end{theorem}
\proof See Appendix \ref{appendix:theorem:1}.\endproof

Fig.~\ref{fig:fig1} plots the cdf curves of $\lambda_{\sf max}$ with different system configurations. Results in this figure demonstrate that the analytical results are in perfect agreement with the Monte Carlo simulation results, which confirms the correctness of the analytical expressions.

The following corollaries present two special cases for which the cdf expression in (\ref{eqn:exactcdf}) can be reduced to a simple form. The results will be useful to study the outage probability of the systems where one of the nodes is equipped with only a single antenna.

\begin{corollary}\label{coro:ns1}
When $n_s=1$, (\ref{eqn:exactcdf}) reduces to
\begin{equation}
{\cal F}_{\lambda_{\sf max}}(x)=\frac{(-1)^{q-1}\sum_{j=1}^{q}D_j\left(\sum_{l=0}^{j-1}{j-1\choose l}a^l\left(\Gamma(\eta)-2e^{-ax}x^{\frac{\eta}{2}}K_{\eta}(2\sqrt{x})\right)\right)}{\prod_{i=1}^{q}\Gamma(q-i+1)\Gamma(p-i+1)},
\end{equation}
where $\eta \triangleq p+q+l-j$, and $D_j$ is the $(q,j)$-th cofactor of a $q\times q$ matrix whose $(i,j)$-th entry is given by $(-1)^{q-1-i}\sum_{l=0}^{q-i+j-1}{q-i+j-1\choose l}a^l\Gamma(p+l+i-j)$.
\end{corollary}
\proof When $n_s = 1$, we have $s=1$ and $t = q$. As a consequence, the desired result follows by substituting $s=1$ into (\ref{eqn:exactcdf}) and then applying Laplace's expansion along the last row of the determinants, which completes the proof.\endproof

\begin{corollary}\label{coro:q1}
When $q=1$, (\ref{eqn:exactcdf}) reduces to
\begin{equation}
{\cal F}_{\lambda_{\sf max}}(x) = 1-\frac{2e^{-ax}}{\Gamma(p)}\sum_{k=0}^{n_s-1}\frac{1}{k!}\sum_{l=0}^{k}{k\choose l}a^lx^{\frac{p+l+k}{2}}K_{p+l-k}(2\sqrt{x}).
\end{equation}
\end{corollary}

\proof The proof is straightforward and thus omitted.\endproof

\subsection{Exact Expressions for the pdf}

The following theorem presents the pdf of the maximum eigenvalue $\lambda_{\sf max}$. This will be used to compute the ergodic capacity of optimal beamforming scheme in MIMO dual-hop AF systems.

\begin{theorem}\label{theorem:pdf}
Let $ {\bf{H}}_1  \in \mathcal{C}^{n_r  \times n_s}$ and $ {\bf{H}}_2  \in \mathcal{C}^{n_d  \times n_r}$, and $a$ being a non-negative constant, the pdf of the maximum eigenvalue $\lambda_{\sf max}$ of $\mathbf{H}_1^{\dag}\mathbf{H}_2^{\dag}\left(a\mathbf{H}_2\mathbf{H}_2^{\dag}+\mathbf{I}_{n_d}\right)^{-1}\mathbf{H}_2\mathbf{H}_1$ is given by
\begin{align}\label{eqn:exactpdf}
f_{\lambda_{\sf max}}(x) =\frac{(-1)^{n_s(t-n_s)}\sum_{l=q-s+1}^{q}\det\left(\mathbf{\Phi}_l(x)\right)}{\prod_{i=1}^{q}\Gamma(q-i+1)\Gamma(p-i+1)},
\end{align}
where $\mathbf{\Phi}_l(x)$ is a $q\times q$ matrix with entries
\begin{equation}\label{eqn:defphi}
[\mathbf{\Phi}_l(x)]_{i,j}=\left\{\begin{array}{cc}
[\mathbf{\Phi}(x)]_{i,j} & \mbox{for }i\neq l,\\
2e^{-ax}\sum_{k=0}^{t-i}\frac{1}{k!}\sum_{l=0}^{\tau(i,j)+k}{\tau(i,j)+k\choose l}a^lg(x)& \mbox{for }i=l.
\end{array}\right.
\end{equation}
where
\begin{align}
g(x) = (ax-k)x^{\frac{\theta(i,j)+l+k-1}{2}}K_{\theta(i,j)+l-k+1}(2\sqrt{x})+x^{\frac{\theta(i,j)+l+k}{2}}K_{\theta(i,j)+l-k}(2\sqrt{x}).
\end{align}
\end{theorem}
\proof See Appendix \ref{appendix:theorem:pdf}.\endproof

Similar to our cdf analysis before, the following corollaries present two special cases for which the pdf expression in (\ref{eqn:exactpdf}) is simplified. The results will be used to study the ergodic capacity of the systems where one of the nodes is equipped with only a single antenna.

\begin{corollary}\label{coro:pdfns1}
When $n_s=1$, (\ref{eqn:exactpdf}) reduces to
\begin{align}
f_{\lambda_{\sf max}}(x) =\frac{(-1)^{q-1}2\sum_{j=1}^{q}D_je^{-ax}\left(\sum_{l=0}^{j-1}{j-1\choose
l}a^l\left(ax^{\frac{\eta}{2}}K_{\eta}(2\sqrt{x})+x^{\frac{\eta-1}{2}}K_{\eta-1}(2\sqrt{x})\right)\right)}{\prod_{i=1}^{q}\Gamma(q-i+1)\Gamma(p-i+1)},
\end{align}
where $\eta$ and $D_j$ are defined in Corollary \ref{coro:ns1}.
\end{corollary}
\proof When $n_s = 1$, we have $s=1$ and $t = q$. Hence, the desired result is obtained by substituting $s=1$ into (\ref{eqn:exactpdf}) and using Laplace's expansion on the last row of the determinants.\endproof

\begin{corollary}\label{coro:pdfq1}
When $q=1$, (\ref{eqn:exactpdf}) reduces to
\begin{align}
f_{\lambda_{\sf max}}(x) = \frac{2e^{-ax}}{\Gamma(p)}\sum_{k=0}^{n_s-1}\frac{x^k}{k!}\sum_{l=0}^k{k\choose l}a^{k-l}\left(\left(a-\frac{k}{x}\right)x^{\frac{p-l}{2}}K_{p-l}(2\sqrt{x})+x^{\frac{p-l-1}{2}}K_{p-l-1}(2\sqrt{x})\right).
\end{align}
\end{corollary}
\proof The proof is straightforward and thus omitted.\endproof

\subsection{Asymptotic Expansions for the cdf and pdf}
Here, we present some asymptotic expansions for the cdf and pdf of $\lambda_{\sf max}$. Although obtaining such expansions for general systems with arbitrary $n_s$, $n_r$ and $n_d$ is a challenging task, simple expansions are possible for the important special cases where one of the nodes has only one antenna, as we show in the following. These results will be used for investigating the asymptotic outage probability and for deriving the diversity order and array gain of the systems.

\begin{theorem}\label{theorem:asy2}
When $n_s = 1$, the cdf and pdf of $\lambda_{\sf max}$ have the asymptotic expansions:
\begin{align}
{\cal F}_{\lambda_{\sf max}}(x) &= \frac{v_1}{q}x^q+o(x^{q}),\\
{f}_{\lambda_{\sf max}}(x) & = v_1x^{q-1}+o(x^{q-1}),
\end{align}
where
\begin{equation}
v_1 = \frac{\det({\bf \Psi})}{\Gamma(q)\prod_{i=1}^{q}\Gamma(p-i+1)\Gamma(q-i+1)},
\end{equation}
where ${\bf \Psi}$ is a $q\times q$ matrix with entries
\begin{equation}
[{\bf \Psi}]_{i,j} = \sum_{k=0}^{2q-i-j+1}{2q-i-j+1\choose k}a^k\Gamma(p-q+k+i+j-2).
\end{equation}
\end{theorem}
\proof See Appendix \ref{appendix:theorem:asy2}.\endproof

\begin{theorem}\label{theorem:asy1}
When $q = 1$, the cdf and pdf of $\lambda_{\sf max}$ have the asymptotic expansions:
\begin{align}
{\cal F}_{\lambda_{\sf max}}(x) &= \frac{v_2}{m}x^m+o(x^m),\\
{f}_{\lambda_{\sf max}}(x) & = v_2x^{m-1}+o(x^{m-1}),
\end{align}
where
\begin{equation}
v_2 = \left\{\begin{array}{cc}
\frac{\sum_{i=0}^{n_s}{n_s\choose i}a^i \Gamma(p-n_s+i)}{\Gamma(p)\Gamma(n_s)}&\mbox{\quad if\quad} p > n_s,\\
\frac{-c-\ln x +\sum_{i=1}^{n_s}{n_s\choose i}a^i \Gamma(i)}{\Gamma(p)\Gamma(n_s)}&\mbox{\quad if\quad} p = n_s,\\
\frac{\Gamma(n_s-p)}{\Gamma(p)\Gamma(n_s)} &\mbox{\quad if\quad} p < n_s.\\
\end{array}\right.
\end{equation}
\end{theorem}
\proof See Appendix \ref{appendix:theorem:asy}.\endproof

\subsection{Exact Expressions for the Moments}
In this section, we present exact expressions for the moments of $\lambda_{\sf max}$ for the important special cases corresponding to the systems where one of the nodes is equipped with a single antenna. These results will be used for investigating the ergodic capacity of the system.
\begin{theorem}\label{theorem:moment1}
When $n_s =1$, the $m$-th moment of $\lambda_{\sf max}$ is given by
\begin{multline}
{\tt E}\{(\lambda_{\sf max})^m\}= \frac{(-1)^{q-1}\Gamma(m+1)}{\prod_{i=1}^{q}\Gamma(q-i+1)\Gamma(p-i+1)}
\sum_{j=1}^{q}D_j\\\left(\sum_{l=0}^{j-1}{j-1\choose
l}a^{l-m}\Gamma(\eta+m)\left((\eta+m)U\left(m+1,1-\eta,\frac{1}{a}\right)+a^{-1}U\left(m+1,2-\eta,\frac{1}{a}\right)\right)\right).
\end{multline}
\end{theorem}
\proof Starting from the definition of the moment, the desired result can be obtained with the help of the following
integration relationship
\begin{equation}\label{eqn:integral}
\int_0^{\infty}x^{\mu}e^{- mx}K_v(2\sqrt{\beta x})dx
=\frac{\beta^{-\frac{v}{2}}\Gamma\left(\mu+\frac{v}{2}+1\right)\Gamma\left(\mu-\frac{v}{2}+1\right)}{2m^{\mu-\frac{v}{2}+1}
}U\left(\mu-\frac{v}{2}+1,1-v,\frac{\beta}{m}\right).
\end{equation} \endproof

\begin{theorem}\label{theorem:moment2}
When $q =1$, the $m$-th moment of $\lambda_{\sf max}$ is given by
\begin{equation}
{\tt E}\{(\lambda_{\sf max})^m\}= \frac{\Gamma(n_s+m)}{\Gamma(p)\Gamma(n_s)}\sum_{i=0}^{n_s}{n_s\choose i}\frac{\Gamma(p+i+m)}{a^{n_s+m-i}}U\left(n_s+m,1-p+n_s-i,\frac{1}{a}\right).
\end{equation}
\end{theorem}
\proof The proof is similar to that of Theorem \ref{theorem:moment1}.\endproof

\subsection{Asymptotic Approximations for the cdf}
In this section, we present the asymptotic approximation for the cdf of $\lambda_{\sf max}$ in the case where the number of antennas at one of the nodes becomes large. These results are useful when studying the average SER and ergodic capacity of the system.
\begin{theorem}\label{theorem:asyantenna1}
When the number of antennas at one of the nodes becomes large, $\lambda_{\sf max}$ is statistically equivalent to
\begin{equation}
\lambda_{\sf max} = \left\{\begin{array}{cc}
\frac{n_s\lambda_{\sf max1}}{a\lambda_{\sf max1}+1} & \mbox{\quad if \quad} n_s\rightarrow\infty,\\
\frac{n_d}{an_d+1}\lambda_{\sf max2}&\mbox{\quad if \quad} n_d\rightarrow\infty,\\
\frac{n_r}{an_r+1}\lambda_{\sf max3} &\mbox{\quad if \quad} n_r\rightarrow\infty,
\end{array}\right.
\end{equation}
where $\lambda_{\sf max1}$, $\lambda_{\sf max2}$ and $\lambda_{\sf max3}$ are the maximum eigenvalues of the channel matrices ${\bf H}_2^{\dag}{\bf H}_2$, ${{\bf H}}_1^{\dag} {{\bf H}}_1$, and $\bar{{\bf H}}_1^{\dag}\bar {{\bf H}}_1$ with $\bar{\bf H}_1\in {\cal CN}^{n_s\times n_d}$, respectively.
\end{theorem}
\proof See Appendix \ref{appendix:theorem:asyantenna1}.\endproof

\section{Performance Analysis}
In this section, we use the new statistical results of $\lambda_{\sf max}$ derived in the previous section to investigate three important performance measures for beamforming schemes in MIMO dual-hop AF systems, i.e., outage probability, average SER and ergodic capacity.

\subsection{Outage Probability}
The outage probability is an important quality of service measure, defined as the probability that the instantaneous SNR falls below a pre-defined threshold $\gamma_{\sf th}$, and can be expressed as
\begin{align}
P_{\sf out} (\gamma_{\sf th})& = {\sf Pr}(\gamma <\gamma_{\sf th}) = {\sf Pr}\left(\lambda_{\sf max} <\frac{\gamma_{\sf
th}}{a\rho}\right)=\frac{(-1)^{n_s(t-n_s)}\det\left(\mathbf{\Phi}\left(\frac{\gamma_{\sf
th}}{a\rho}\right)\right)}{\prod_{i=1}^{q}\Gamma(q-i+1)\Gamma(p-i+1)},\label{eqn:outagedef}
\end{align}
Note that (\ref{eqn:outagedef}) presents the exact outage probability expression and allows for efficient evaluation of
the outage probability of the optimal beamforming MIMO dual-hop AF systems for arbitrary numbers of antennas at the
source, relay and destination terminals, at any desired SNR.

Fig.~\ref{fig:fig21} investigates the joint impact of $n_s$, $n_r$, and $n_d$ on the system outage probability. Results indicate that while it is always beneficial to have more antennas at the transmitter than the receiver, the advantage of doing so heavily relies on $n_r$. For small $n_r$, e.g., $n_r =1$, the benefit of putting more antennas at the transmitter is considerable. However, when $n_r$ becomes large, e.g., $n_r =10$, such advantage is significantly reduced, and $n_s$ and $n_d$ play a rather symmetric role.

We now examine in more detail the outage probability for the special cases where the number of antennas at one of the nodes is one. We first consider the case $n_s=1$, i.e., when the number of antennas at source node is one. From (\ref{eqn:outagedef}) and Corollary \ref{coro:ns1}, the outage probability is
\begin{equation}
P_{\sf out}(\gamma_{\sf th})=\frac{(-1)^{q-1}\sum_{j=1}^{q}D_j\left(\sum_{l=0}^{j-1}{j-1\choose l}a^l\left(\Gamma(\eta)-2e^{-\frac{\gamma_{\sf th}}{\rho}}\left(\frac{\gamma_{\sf
th}}{a\rho}\right)^{\frac{\eta}{2}}K_{\eta}(2\sqrt{\frac{\gamma_{\sf th}}{a\rho}})\right)\right)}{\prod_{i=1}^{q}\Gamma(q-i+1)\Gamma(p-i+1)}.
\end{equation}

To gain more insights, it is of great interest to look into the low outage regime (or the high SNR regime). The outage performance in this regime can be generally characterized by two key parameters, i.e., diversity order and array gain. The diversity order is defined as the slope of the outage probability curve versus SNR curve (plotted on log-log scale), as SNR grows large. It is worth pointing out in the relaying system that high SNR consideration implies that both the source and relay power gains increase simultaneously.\footnote{If either of the source or relay power gain is fixed, then the outage probability of the system reaches an error floor when the other power gain increases. Therefore, the diversity order achieved in such scenario is zero.} Therefore, we are interested in the scenario where both $\rho$ and $\alpha$ increase to infinity with a fixed ratio, i.e., $\alpha = k\rho$ and $\rho\rightarrow \infty$.

At high SNR, from Theorem \ref{theorem:asy2}, the outage probability can be approximated as
\begin{equation}
P_{\sf out}(\gamma_{\sf th})\approx \frac{\det({\bf \Psi})}{\Gamma(q+1)\prod_{i=1}^{q}\Gamma(p-i+1)\Gamma(q-i+1)}\left(\frac{\gamma_{\sf th}}{a\rho}\right)^q,
\end{equation}
which shows that the system achieves diversity order of $q$.

Now, we consider the case when $q=1$, i.e., the number of antennas at relay or destination node is one. From (\ref{eqn:outagedef}) and Corollary \ref{coro:q1}, the outage probability of the system can be expressed as
\begin{equation}\label{eqn:31}
P_{\sf out}(\gamma_{\sf th}) = 1-\frac{2e^{-\frac{\gamma_{\sf th}}{\rho}}}{\Gamma(p)}\sum_{k=0}^{n_s-1}\frac{1}{k!}\sum_{l=0}^{k}{k\choose l}a^l\left(\frac{\gamma_{\sf th}}{a\rho}\right)^{\frac{p+l+k}{2}}K_{p+l-k}\left(2\sqrt{\frac{\gamma_{\sf th}}{a\rho}}\right).
\end{equation}
Note, when $n_s=1$ and $p=1$, Eq. (\ref{eqn:31}) reduces to the
results presented in \cite[Eq. (9)]{M.Hasna}. At high SNR, from
Theorem \ref{theorem:asy1}, the outage probability can be
approximated as
\begin{align}
P_{\sf out}(\gamma_{\sf th}) \approx \left\{\begin{array}{cc}
\frac{n_r^m\sum_{i=0}^{n_s}{n_s\choose i}\left(\frac{k}{n_r}\right)^i \Gamma(p-n_s+i)}{mk^m\Gamma(p)\Gamma(n_s)}\left(\frac{\gamma_{\sf th}}{\rho}\right)^m&\mbox{\quad if\quad} p > n_s,\\
\frac{n_r^m\ln \left(\frac{k\rho}{n_r\gamma_{\sf th}}\right) }{mk^m\Gamma(p)\Gamma(n_s)}\left(\frac{\gamma_{\sf th}}{\rho}\right)^m&\mbox{\quad if\quad} p = n_s,\\
\frac{n_r^m\Gamma(n_s-p)}{mk^m\Gamma(p)\Gamma(n_s)} \left(\frac{\gamma_{\sf th}}{\rho}\right)^m&\mbox{\quad if\quad} p < n_s,
\end{array}\right.\notag
\end{align}
which shows that diversity order of $m$ is achieved by the system. To see how the outage probability behaves when $n_s$ increases, we consider the case $p<n_s$. When $n_s$ is increased by one, the outage probability is reduced by a factor of $\frac{p}{n_s}$, which indicates that increasing the number of antennas at the source is always beneficial, but the benefit diminishes as $n_s$ becomes large.

Fig.~\ref{fig:fig2} shows the outage probability of the system with different $n_d$ when $n_s = 3$, $n_r=1$. We observe that the Monte Carlo simulation results are in prefect agreement with our exact analytical results. Also, we can see that the high SNR approximation results are quite accurate even at moderate SNRs, i.e., $\rho = 20 \mbox{dB}$. Moreover, the outage curves suggest that diversity order of $2$ is achieved for $n_d=2$, and $3$ is achieved for $n_d = 3, 4$, as expected.

\subsection{SER}
In addition to the outage probability, average SER is also another important metric for characterizing the performance of a communication system. For most modulation formats, the average SER can be evaluated as \cite{Proakis}
\begin{equation}\label{eqn:ser}
\mbox{SER}={\tt E}_{\gamma}\left[a_{1} Q\left(\sqrt{2a_2\gamma}\right)\right],
\end{equation}
where $Q(\cdot)$ is the Gaussian Q-function, and $a_1$, $a_2$ are modulation-specific constants.

Using (\ref{eqn:ser}) and the pdf of $\lambda_{\sf max}$ in Theorem \ref{theorem:pdf}, the average SER can be evaluated as
\begin{equation}\label{eqn:sergeneral}
\mbox{SER}_{\sf BF}(\rho) =\frac{(-1)^{n_s(t-n_s)}\sum_{l=q-s+1}^{q}\int_0^{\infty}a_1 Q\left(\sqrt{2a_2 a\rho x}\right)\det\left(\mathbf{\Phi}_l(x)\right)dx}{\prod_{i=1}^{q}\Gamma(q-i+1)\Gamma(p-i+1)}.
\end{equation}
With arbitrary numbers of antennas at the source, relay and destination terminals, it does not appear that the integral in (\ref{eqn:sergeneral}) can be expressed in closed-form. However, numerical integration can be performed to calculate SER much more efficiently than is possible via Monte Carlo simulations.

For the special cases where the number of antennas at one of the nodes is large, accurate approximations can be obtained using Theorem \ref{theorem:asyantenna1} and the results in \cite{M.McKay}. Define ${\cal S}(c_1, c_2, \bar{\rho})$\footnote{The closed-form expression of ${\cal S}(c_1, c_2, \bar{\rho})$ is given in \cite{M.McKay}. To avoid cumbersome notations, we do not give the explicit expression here. It is also worth noting that the average SNR here is slightly different to that in \cite{M.McKay}, and it is given by $\bar{\rho} = a\rho$.} as the average SER of optimal beamforming in uncorrelated Rayleigh fading channels with $c_1$ transmit antennas and $c_2$ receive antennas at average SNR $\bar{\rho}$. Then, we have the following result.
\begin{proposition}\label{prop:ser}
The average SER of optimal beamforming systems in MIMO dual-hop channels can be approximated as
\begin{equation}
\mbox{SER}_{\sf BF}(\rho) \approx \left\{\begin{array}{cc}
{\cal S}\left(n_d, n_s, \frac{n_r}{an_r+1}\bar{\rho}\right)& \mbox{if\quad}n_r \rightarrow \infty,\\
{\cal S}\left(n_r, n_s, \frac{n_d}{an_d+1}\bar{\rho}\right) &  \mbox{if\quad}n_d \rightarrow \infty.
\end{array}\right.
\end{equation}
\end{proposition}

Proposition \ref{prop:ser} offers an efficient means to evaluate the average SER of the system, and as illustrated in Fig.~\ref{fig:fig3}, the asymptotic approximations are very accurate even at reasonably small number of antennas, i.e., $n_r$ or  $n_d = 10$, and it performs particularly good in the low SNR regime, where it is observed that the approximation curves and Monte Carlo simulation curves almost overlap. Moreover, the numerical results also suggest that for a given number of transmit antennas, it is beneficial to deploy more antennas at the receiver than at the relay.

It is well known that optimal beamforming scheme achieves full diversity of MIMO systems, i.e., diversity of $c_1c_2$ is achieved for MIMO systems with $c_1$ transmit antennas and $c_2$ receive antennas. Hence, another important implication of Theorem \ref{theorem:asyantenna1} and Proposition \ref{prop:ser} is that full diversity order of $\frac{n_sn_rn_d}{n}$ is achieved by optimal beamforming in MIMO dual-hop AF systems when the number of antennas at one of the nodes becomes large. Now, we consider the special cases, where one of the nodes has only one antenna, and derive exact closed-form expressions for the average SER.

\begin{proposition}\label{prop:1}
When $n_s = 1$, the average SER of beamforming schemes in MIMO dual-hop AF systems can be expressed as
\begin{multline}
\mbox{SER}_{\sf BF}(\rho) =\frac{a_1\sqrt{a_2}(-1)^{q-1}}{2\sqrt{\pi}\prod_{i=1}^{q}\Gamma(q-i+1)\Gamma(p-i+1)}\\
\sum_{j=1}^{q}D_j\left(\sum_{l=0}^{j-1}{j-1\choose l}a^l\left(\Gamma(\eta)\sqrt{\frac{\pi}{a_2}}-\frac{\Gamma(\eta+1/2)\Gamma(1/2)}{(1/\rho+a_2)^{1/2}}U\left(\frac{1}{2},1-\eta,\frac{1}{a(\rho a_2+1)}\right)\right)\right).
\end{multline}
\end{proposition}
\proof The average SER of the system can be alternatively computed via
\begin{equation}
\mbox{SER}_{\sf BF}(\rho) = \frac{\alpha\sqrt{\beta}}{2\sqrt{\pi}}\int_0^{\infty}\frac{e^{-\beta x}}{\sqrt{x}}{\cal F}_{\gamma}(x)dx.
\end{equation}
Substituting the cdf expression presented in Corollary \ref{coro:ns1}, the desired result can be obtained with the help
of the integration relationship Eq. (\ref{eqn:integral}), and the fact that $\int_0^{\infty}\frac{e^{-\beta
x}}{\sqrt{x}}dx =\sqrt{\frac{\pi}{\beta}}$.
\endproof

To derive the diversity order and array gain of optimal beamforming in MIMO dual-hop systems, we now analyze the SER performance in the high SNR regime. To this end, with the aid of Theorem \ref{theorem:asy1}, we can invoke a general parameterized SER result from \cite{Z.Wang} and perform some basic algebraic manipulations to obtain the following high SNR SER expression:
\begin{equation}
\mbox{SER}_{\sf BF}(\rho)\approx \frac{a_1 n_r^{q}\Gamma(n_s+1/2)\det({\bf \Psi})}{2\sqrt{\pi}(a_2k)^q\Gamma(q+1)\prod_{i=1}^{q}\Gamma(p-i+1)\Gamma(q-i+1)}\rho^{-q},
\end{equation}
which indicates that the system achieves diversity order of $q$, as expected.

\begin{proposition}\label{prop:2}
When $q = 1$, the average SER of beamforming schemes in MIMO dual-hop AF systems can be expressed as
\begin{multline}
\mbox{SER}_{\sf BF}(\rho) =
\frac{a_1}{2}-\frac{a_1\sqrt{a_2}}{2\sqrt{\pi}\Gamma(\rho)}\sum_{k=0}^{n_s-1}\frac{1}{k!}\sum_{l=0}^{k}{k\choose
l}a^l\frac{\Gamma\left(p+l+\frac{1}{2}\right)\Gamma\left(k+\frac{1}{2}\right)}{(a\rho)^k(a_2+1/\rho)^{k+1/2}}\\
\times U\left(k+\frac{1}{2},1+k-p-l,\frac{1}{a(\rho a_2+1)}\right).
\end{multline}
\end{proposition}
\proof The proof is straightforward and omitted.\endproof

Similarly, the average SER at high SNR can be approximated as
\begin{equation}\label{eqn:ser:high}
\mbox{SER}_{\sf BF}(\rho) = \left\{\begin{array}{cc}
\frac{a_1 n_r^{n_s}\Gamma(n_s+1/2)\sum_{i=0}^{n_s}{n_s\choose i}a^i \Gamma(p-n_s+i)}{2\sqrt{\pi} (ka_2)^{n_s}\Gamma(p)\Gamma(n_s+1)}\rho^{-n_s}&\mbox{\quad if\quad} p > n_s,\\
\frac{a_1n_r^{p}\Gamma(p+1/2)\ln  \left(\frac{k\rho}{n_r}\right)}{2\sqrt{\pi}(ka_2)^{p} \Gamma(p+1)\Gamma(n_s)}\rho^{-p}&\mbox{\quad if\quad} p = n_s,\\
 \frac{a_1n_r^{p}\Gamma(p+1/2)\Gamma(n_s-p)}{2\sqrt{\pi} (ka_2)^{p}\Gamma(p+1)\Gamma(n_s)}\rho^{-p} &\mbox{\quad if\quad} p < n_s,\\
\end{array}\right.
\end{equation}
which illustrates that diversity order of $m$ is achieved. It is also of interest to compare the SER performance between optimal beamforming and OSTBC. In \cite{Pratha}, the high SNR approximation of SER was derived for the OSTBC system employing BPSK modulation as
\begin{equation}\label{eqn:ser:ostbc}
\mbox{SER}_{\sf OSTBC}(\rho) = \left\{\begin{array}{cc}
\frac{(Rn_s)^{n_s}\Gamma\left(n_s+\frac{1}{2}\right)\sum_{i=0}^{n_s}{n_s\choose i}a^i \Gamma(p-n_s+i)\left(\frac{k}{n_r}\right)^{-l}}{2\sqrt{\pi} \Gamma(p)\Gamma(n_s+1)}\rho^{-n_s}&\mbox{\quad if\quad} p > n_s,\\
\frac{(Rn_sn_r)^{p}\Gamma\left(p+\frac{1}{2}\right)\Gamma(n_s-p)}{2\sqrt{\pi} k^{p}\Gamma(p+1)\Gamma(n_s)}\rho^{-p} &\mbox{\quad if\quad} p < n_s,
\end{array}\right.
\end{equation}
where $R$ denotes the code rate of OSTBC schemes. Comparing (\ref{eqn:ser:high}) and (\ref{eqn:ser:ostbc}), it can be seen that both optimal beamforming and OSTBC achieve the same diversity order $m$. Since beamforming scheme exploits channel knowledge at the transmitter while OSTBC dose not, one would therefore expect beamforming scheme to outperform OSTBC. To see this, let us assume $\mbox{SER}_{\sf BF}(\rho_{\sf BF}) = \mbox{SER}_{\sf OSTBC} (\rho_{\sf OSTBC})$, and quantify the advantage of beamforming scheme in terms of SNRs. For the cases $p\neq n_s$, it is not difficult to observe that
\begin{align}
\Delta_{\sf dB} =(\rho_{\sf OSTBC})_{\sf dB}-(\rho_{\sf BF})_{\sf dB} = (Rn_s)_{\sf dB},
\end{align}
which indicates a significant power saving of beamforming schemes over OSTBC schemes.

Fig.~\ref{fig:fig4} compares the average SER performance of optimal beamforming and OSTBC. For the OSTBC simulations, a full-rate Alamouti code \cite{Alamouti} was used, so that $R =1$, and the analytical result was plotted as per \cite[(26)]{Pratha}.\footnote{There is a typo in the expression in \cite{Pratha}: the second argument of the Hypergeometric U-function should be ``$p+l-k+1$'' instead of ``$p+l-k-1$''.} Clearly, we see that optimal beamforming and OSTBC achieve the same diversity order of $2$. Moreover, result indicate that the optimal beamforming scheme outperforms OSTBC by $3\mbox{dB}$ in terms of SNR, as is expected.

\subsection{Ergodic Capacity}
In this section, we study the ergodic capacity achieved by the optimal beamforming scheme. The exact ergodic capacity of the system can be expressed as
\begin{equation}\label{eqn:cap}
C(\rho) =\frac{1}{2}{\tt E}_{\lambda_{\sf max}}\left\{\log_2(1+a\rho \lambda_{\sf max})\right\}.
\end{equation}
Using (\ref{eqn:cap}) and the pdf of $\lambda_{\sf max}$ in Theorem \ref{theorem:pdf}, the average SER can be evaluated by
\begin{equation}\label{eqn:capgeneral}
C(\rho) =\frac{(-1)^{n_s(t-n_s)}\sum_{l=q-s+1}^{q}\int_0^{\infty}\log_2(1+a\rho \lambda_{\sf max})\det\left(\mathbf{\Phi}_l(x)\right)dx}{2\prod_{i=1}^{q}\Gamma(q-i+1)\Gamma(p-i+1)}.
\end{equation}
With arbitrary numbers of antennas at the source, relay and destination terminals, the integral in (\ref{eqn:capgeneral}) is not known to be evaluated in closed-form. However, numerical integration can be performed to calculate the ergodic capacity much more efficiently than is possible via Monte Carlo simulations. For the special cases where the number of antennas at one of the nodes becomes large, very accurate approximations can be obtained.

Let ${\cal I}(c_1, c_2, \bar{\rho})$\footnote{The closed-form expression of ${\cal I}(c_1, c_2, \bar{\rho})$ is given in \cite{M.McKay}.} denote the exact ergodic capacity achieved by optimal beamforming in uncorrelated Rayleigh fading channels with $c_1$ transmit antennas and $c_2$ receive antennas at average SNR $\bar{\rho}$. Then, we have the following result.
\begin{proposition}\label{prop:appcap}
The ergodic capacity achieved by optimal beamforming in MIMO dual-hop channels can be approximated as
\begin{equation}
C(\rho) \approx \left\{\begin{array}{cc}
\frac{1}{2}\left({\cal I}\left(n_r, n_d, n_s\bar{\rho}+a\right)-{\cal I}\left(n_r, n_d, a\right)\right) & \mbox{as }n_s \rightarrow \infty,\\
\frac{1}{2} {\cal I}\left(n_d, n_s, \frac{n_r}{an_r+1}\bar{\rho}\right)& \mbox{as }n_r \rightarrow \infty,\\
\frac{1}{2} {\cal I}\left(n_r, n_s, \frac{n_d}{an_d+1}\bar{\rho}\right) & \mbox{as }n_d \rightarrow \infty.
\end{array}\right.
\end{equation}
\end{proposition}
\proof The results for the last two cases can be obtained via straightforward application of Theorem
\ref{theorem:asyantenna1}. For the first case, the key observation is that the ergodic capacity of the system can be
approximated as
\begin{align}
C(\rho) & =\frac{1}{2}{\tt E}\left\{\log_2\left(1+(n_s\bar{\rho}+a)\lambda_{\sf max1}\right)\right\}- \frac{1}{2}{\tt
E}\left\{\log_2\left(1+a\lambda_{\sf max1}\right)\right\}.
\end{align}
Then, the desired result follows immediately.
\endproof

Fig.~\ref{fig:fig5} illustrates the ergodic capacity of optimal beamforming MIMO dual-hop AF systems. As we can see, the capacity approximations in Proposition \ref{prop:appcap} are accurate over the entire SNR range of interest with even moderate number of antennas. Results also show that employing more antennas at the source results in better capacity performance than at the receiver, which agrees with the observations made from the outage probability and SER analysis.

The above proposition deals with the case where one of the nodes is equipped with a large number of antennas. Next, we consider the case where one of the nodes is equipped with only one antenna. When $n_s=1$ or $q =1$, we can obtain a tight upper bound for the ergodic capacity via Jensen's inequality, i.e.,
\begin{equation}
C(\rho) \leq C_{\sf up} =\frac{1}{2}\log_2(1+a\rho {\tt E}\{\lambda_{\sf max}\}).
\end{equation}
Alternatively, the following simple and accurate capacity approximation can be used \cite{J.Perez}
\begin{equation}
C(\rho) \approx C_{\sf app}=\frac{1}{2}\log_2e\left(\ln(1+a\rho {\sf E}\{\lambda_{\sf max}\})-\frac{a^2\rho^2({\tt E}\{\lambda_{\sf max}^2\}-{\sf E}\{\lambda_{\sf max}\}^2)}{2(1+a\rho{\sf E}\{\lambda_{\sf max}\})^2}\right).
\end{equation}

To obtain further insights, we look into the high SNR regimes, and derive simple and closed-form expressions for the
two key parameters dictating the ergodic capacity, i.e., high SNR slope ${\cal S}_{\infty}$ and power offset ${\cal
L}_{\infty}$. The high SNR slope ${\cal S}_{\infty}$ and the power offset ${\cal L}_{\infty}$ can be obtained via
\begin{equation}
{\cal S}_{\infty} = \lim_{\rho\rightarrow\infty}\frac{C(\rho)}{\log_2\rho},\quad {\cal L}_{\infty} =
\lim_{\rho\rightarrow\infty}\left(\log_2\rho-\frac{C(\rho)}{{\cal S}_{\infty}}\right),
\end{equation}
respectively.
%

From (\ref{eqn:cap}), we can easily establish that ${\cal S}_{\infty} = \frac{1}{2}$. Hence, the power offset ${\cal L}_{\infty}$ can be computed as (the proof is given in Appendix \ref{appendix:eqn:poweroffset})
\begin{equation}\label{eqn:poweroffset}
{\cal L}_{\infty,{\sf BF}}=\log_2\frac{n_r}{k}-\log_2e\left(\psi(n_s)+\psi(p)-\sum_{i=0}^{p-1}e^{\frac{n_r}{k}}E_{i+1}\left(\frac{n_r}{k}\right)\right).
\end{equation}
Note from the above expression that while the impact of $p$ on ${\cal L}_{\infty}$ is not immediately clear, increasing $n_s$ reduces ${\cal L}_{\infty}$, and hence improves the ergodic capacity of the system. More precisely, when the number of transmit antennas is increased by $k_1$, ${\cal L}_{\infty,{\sf BF}}$ will be reduced by $\log_2e(\psi(n_s+k_1)-\psi(n_s))$, which is significant for small $n_s$, but becomes almost negligible when $n_s$ is large.

Fig.~\ref{fig:fig6} illustrates the ergodic capacity of the system with difference antenna configurations. We observe that the capacity upper bound $C_{\sf up}$ performs reasonably well over a wide range of SNR, especially at low SNRs, where $C_{\sf up}$ almost coincides with the exact capacity. In addition, the high SNR approximation also provides a good reference to the exact result at even moderate SNRs, i.e., $5\mbox{dB}$. Moreover, as $n_s$ increases, the performance of both $C_{\sf up}$ and high SNR improve. For instance, the upper bound $C_{\sf up}$, high SNR approximation and the exact result $C$ almost overlap when $n_s = 10$. Furthermore, we see that the capacity approximation $C_{\sf app}$ is incredibly accurate for the entire SNR range, regardless of the numbers of transmit antennas.

\section{Conclusion}
This paper examined the analytical performance of the optimal beamforming scheme in MIMO dual-hop AF systems. After deriving a host of new statistical properties of the maximum eigenvalue, exact and approximate closed-form expressions were derived for three important performance metrics: outage probability, average SER and ergodic capacity, which can be efficiently evaluated. Simple and informative expressions were also obtained for some special cases at high SNR, which illustrate the diversity order and array gains achieved by optimal beamforming. Our results have indicated the coupling effect of the numbers of antennas on the system performance. Moreover, optimal beamforming was shown to achieve the maximum diversity order of MIMO dual-hop AF systems as OSTBC schemes while providing a significant power gain.

\section{Acknowledgement}
The first author would like to thank Dr. Matthew R. McKay for providing the Matlab codes used to evaluate the average SER ${\cal S}(c_1,c_2,\rho)$ and ergodic capacity ${\cal I}(c_1,c_2,\rho)$ of optimal beamforming systems in uncorrelated Rayleigh fading channels.

\appendices

\section{Proof of Theorem \ref{theorem:1}}\label{appendix:theorem:1}
We first present the following lemma, which will be used in the main proof of the theorem.
\begin{lemma}\label{lemma:1}
The statistical property of the maximum eigenvalue of $\mathbf{H}_1^{\dag}\mathbf{H}_2^{\dag}\left(a\mathbf{H}_2\mathbf{H}_2^{\dag}+\mathbf{I}_{n_d}\right)^{-1}\mathbf{H}_2\mathbf{H}_1$ is the same as that of the maximum eigenvalue of $\tilde{\mathbf{H}}_1^{\dag}\mathbf{L}\tilde{ \mathbf{H}}_1$, where $\tilde{\mathbf{H}}_1 \sim {\cal CN}_{q, n_s}(\mathbf{0}, \mathbf{I}_{q}\otimes \mathbf{I}_{n_s})$ and $\mathbf{L} ={\sf diag}\{\beta_1,\dots,\beta_q\}$ with $\beta_i = \frac{\lambda_i^2}{a\lambda_i^2+1}$, in which $\lambda_i$ is the singular value of $\mathbf{H}_2$.
\end{lemma}
\proof We first apply singular value decomposition on $\mathbf{H}_2$. Then, using the unitary invariant property of Gaussian random variables and \cite[Lemma 1]{C.Zhong}, we get the desired result.
\endproof

Now, we find it useful to consider two separate cases, namely: $n_s \geq q$ and $n_s< q$.

When $n_s\geq q$, with the help of Lemma \ref{lemma:1}, the cdf of the maximum eigenvalue $\lambda_{\sf max}$
conditioned on $\mathbf{L}$ was given in \cite{M.Kang} as
\begin{equation}\label{Equation:1}
\mathcal{F}_{\lambda_{\sf max}}(x|\mathbf{L}) = \frac{\det(\mathbf{\Psi}_1(x))}{\det({\bf
V}_1)\prod_{i=1}^{q}\Gamma(n_s-i+1)},
\end{equation}
where $\mathbf{V}_1$ is a $q\times q$ matrix, with the determinant of
\begin{equation}\label{Equation:2}
\det(\mathbf{V}_1)=\left(\prod_{i=1}^{q}\beta_i^{n_s}\right)\prod_{1\leq l \leq k \leq
q}\left(\frac{1}{\beta_k}-\frac{1}{\beta_l}\right),
\end{equation}
and $\mathbf{\Psi}_1(x)$ is a $q\times q$ matrix with entries given by
\begin{equation}\label{Equation:3}
[\mathbf{\Psi}_1(x)]_{i,j}= \beta_j^{n_s-i+1}\gamma\left(n_s-i+1,\frac{x}{\beta_j}\right),
\end{equation}
where $\gamma(n,x)$ is the lower incomplete gamma function.
To obtain the unconditional cdf of $\lambda_{{\sf max}}$, we must further average (\ref{Equation:1}) over the joint pdf
of $\{0 \leq \beta _1  < \cdots < \beta _q \leq \frac{1}{a}\}$ \cite{S.Jin}
\begin{equation} \label{eq:jointPDFg}
f(\beta_1, \dots, \beta_q) = \frac{1}{\prod\nolimits_{i = 1}^q {\Gamma \left( {q - i + 1} \right)\Gamma \left( {p - i+
1} \right)}}  \prod_{i<j}^q ( \beta_j - \beta_i )^2 \prod_{i=1}^{q} \frac{\beta_i^{p - q}
e^{-\frac{\beta_i}{1-a\beta_i}} }{ (1- a \beta_i)^{p + q} }.
\end{equation}
Therefore, the unconditional cdf can be obtained as
\begin{equation}
{\cal F}_{\lambda_{\sf max}}(x)= {\tt E}_{\mathbf{L}}\{{\cal F}_{\lambda_{\sf max}}(x|\mathbf{L})\}=\frac{{\cal
I}_1}{\prod_{i=1}^{q}\Gamma(n_s-i+1){\Gamma \left( {q - i + 1} \right)\Gamma \left( {p - i + 1} \right)}},
\end{equation}
where
\begin{equation}\label{eqn:apnphi1}
{\cal I}_1= \int_{0 \leq \beta _1  < \cdots < \beta _q \leq \frac{1}{a}}\det(\mathbf{\Psi}_1(x))\prod_{1\leq i\leq
j\leq q}(\beta_i-\beta_j) \frac{\beta_i^{p - n_s-1} e^{-\frac{\beta_i}{1-a \beta_i}} }{ (1- a \beta_i)^{p +
q}}d\beta_1\cdots d\beta_q.
\end{equation}
Utilizing the technique proposed in \cite{M.Chiani}, (\ref{eqn:apnphi1}) can be evaluated as $ {\cal
I}_1=\det(\mathbf{\Phi_1(x)})$, where $\mathbf{\Phi_1(x)}$ is a $q\times q$ matrix with entries defined as
\begin{equation}
[\mathbf{\Phi_1(x)}]_{i,j}=
\int_{0}^\frac{1}{a}\beta^{n_s-i+1}\gamma\left(n_s-i+1,\frac{x}{\beta}\right)\beta^{q-j}\frac{\beta^{p - n_s-1}
e^{-\frac{\beta}{1-a \beta}} }{ (1- a \beta)^{p + q} }d\beta.
\end{equation}
Making a change of variable $y = \frac{\beta}{1-a\beta}$, the above integral can be computed as
\begin{equation}
[\mathbf{\Phi_1(x)}]_{i,j}=
\int_0^{\infty}y^{p+q-i-j}(1+ay)^{i+j-2}e^{-y}\gamma\left(n_s-i+1,\frac{x(1+ay)}{y}\right)dy.
\end{equation}
To this end, utilizing the series representation of incomplete gamma function \cite[Eq. (8.351.1)]{Table},
we have
\begin{equation}
[\mathbf{\Phi_1(x)}]_{i,j}= \Gamma(n_s-i+1)(A_1-e^{-ax}B_1(x)),
\end{equation}
where
\begin{align}
A_1 &= \int_0^{\infty}t^{p+q-i-j}(1+at)^{i+j-2}e^{-t}dt=\sum_{l=0}^{i+j-2}{i+j-2\choose l}a^l\Gamma(p+q+l-i-j+1),
\end{align}
and
\begin{align}
B_1(x) &=\sum_{k=0}^{n_s-i}\frac{x^k}{k!}\int_0^{\infty}t^{p+q-i-j-k}(1+at)^{i+j+k-2}e^{-t-\frac{x}{t}}dt\nonumber\\
& = 2\sum_{k=0}^{n_s-i}\frac{x^k}{k!}\sum_{l=0}^{i+j+k-2}{i+j+k-2 \choose
l}a^lx^{\frac{p+q+l-i-j-k+1}{2}}K_{p+q+l-i-j-k+1}(2\sqrt{x}),
\end{align}
where the final expression is obtained using the integration relationship \cite[Eq. (3.471.9)]{Table}.

Now, we proceed to consider the case $n_s \leq q$. The cdf of the maximum eigenvalue of $\mathbf{F}$ conditioned on $\mathbf{L}$ is given by \cite{M.Kang}
\begin{equation}\label{Equation:11}
\mathcal{F}_{\lambda_{\sf
max}}(x|\mathbf{L})=\frac{(-1)^{n_s(q-n_s)}\det(\mathbf{\Psi}_2(x))}{\det(\mathbf{V}_1)\prod_{i=1}^{n_s}\Gamma(n_s-i+1)}
\end{equation}
where $\mathbf{\Psi}_2(x)$ is a $q\times q$ matrix with entries given by
\begin{equation}\label{Equation:13}
[\mathbf{\Psi}_2(x)]_{i,j}=\left\{\begin{array}{ll}
\left(-\frac{1}{\beta_j}\right)^{q-n_s-i} & \mbox{for $i\leq q-n_s$},\\
\beta_j^{q-i+1}\gamma\left(q-i+1,\frac{x}{\beta_j}\right) & \mbox{for $i>q-n_s$}.
\end{array}\right.
\end{equation}
Hence, the unconditional cdf can be obtained by
\begin{equation}
{\cal F}_{\lambda_{\sf max}}(x)= {\tt E}_{\mathbf{L}}\{{\cal F}_{\lambda_{\sf max}}(x|\mathbf{L})\}=\frac{(-1)^{n_s(q-n_s)}{\cal I}_2}{\prod_{i=1}^{n_s}\Gamma(n_s-i+1)\prod_{i=1}^{q}\Gamma \left( {q - i + 1} \right)\Gamma \left( {p - i + 1} \right)},
\end{equation}
where
\begin{equation}
{\cal I}_2 = \int_{0 \leq \beta _1  < \cdots < \beta _q \leq \frac{1}{a}}\det(\mathbf{\Psi}_2(x))\prod_{1\leq i\leq j\leq q}(\beta_i-\beta_j) \frac{\beta_i^{p - n_s-1}
e^{-\frac{\beta_i}{1-a \beta_i}} }{ (1- a \beta_i)^{p + q} }d\beta_1\cdots d\beta_q,
\end{equation}
which can be computed as ${\cal I}_2=\det(\mathbf{\Phi_2(x)})$, where $\mathbf{\Phi_2(x)}$ is a $q\times q$ matrix with
entries defined as
\begin{equation}\label{Equation:16}
[\mathbf{\Phi}_2(x)]_{i,j}=\left\{\begin{array}{ll}
\int_0^\frac{1}{a}(-1)^{q-n_s-i}\frac{\beta^{p+i-j-1}
e^{-\frac{\beta}{1-a \beta}} }{ (1- a \beta)^{p + q} } d\beta& \mbox{for $i\leq q-n_s$},\\
\int_{0}^\frac{1}{a}\beta^{q-i+1}\gamma\left(q-i+1, \frac{x}{\beta}\right)\beta^{q-j}\frac{\beta^{p - n_s-1}
e^{-\frac{\beta}{1-a \beta}} }{ (1- a \beta)^{p + q} }d\beta& \mbox{for $i>q-n_s$}.
\end{array}\right.
\end{equation}
Following similar lines as in the first case, the integrals can be explicitly solved and we have
\begin{equation}\label{Equation:14}
[\mathbf{\Phi}_2(x)]_{i,j}=\left\{\begin{array}{ll}
(-1)^{q-n_s-i}\sum_{l=0}^{q-i+j-1}{q-i+j-1\choose l}a^l\Gamma(p+i+l-j)& \mbox{for $i\leq q-n_s$},\\
\Gamma(q-i+1)(A_2-e^{-ax}B_2(x))& \mbox{for $i>q-n_s$}.
\end{array}\right.
\end{equation}
where
\begin{equation}
A_2 =\sum_{l=0}^{n_s+i+j-q-2}{n_s+i+j-q-2\choose l}a^l\Gamma(2q+p+l-i-j-n_s+1),
\end{equation}
and
\begin{multline}
B_2(x)= 2\sum_{k=0}^{q-i}\frac{x^k}{k!}\\
\sum_{l=0}^{n_s+i+j+k-q-2}{n_s+i+j+k-q-2 \choose
l}a^lx^{\frac{2q+p+l-n_s-i-j-k+1}{2}}K_{2q+p+l-n_s-i-j-k+1}(2\sqrt{x}).
\end{multline}
As such, the unified expression can be obtained by appropriately choosing the parameters.

\section{Proof of Theorem \ref{theorem:pdf}}\label{appendix:theorem:pdf}
Noting that $f_{\lambda_{\sf max}}(x) = \frac{d{\cal F}_{\lambda_{\sf max}}(x)}{dx}$, and making using of Theorem \ref{theorem:1} and a classical formula for the derivative of a determinant, the pdf of the maximum eigenvalue can be written as
\begin{equation}
f_{\lambda_{\sf max}}(x) =\frac{(-1)^{n_s(t-n_s)}\sum_{l=q-s+1}^{q}\det\left(\mathbf{\Phi}_l(x)\right)}{\prod_{i=1}^{q}\Gamma(q-i+1)\Gamma(p-i+1)},
\end{equation}
where $\mathbf{\Phi}_l(x)$ is a $q\times q$ matrix with $(i,j)$-th entry given by
\begin{equation}
[\mathbf{\Phi}_l(x)]_{i,j}=\left\{\begin{array}{ll}
[\mathbf{\Phi}(x)]_{i,j} & \mbox{for }i\neq l,\\
\frac{d(A-e^{-ax}B(x))}{dx} & \mbox{for }i=l.
\end{array}\right.
\end{equation}
The derivative can be computed as
\begin{equation}
\frac{d(A-e^{-ax}B(x))}{dx} = e^{-ax}\left(aB(x)-\frac{dB(x)}{dx}\right).
\end{equation}
To proceed, we compute the first derivative of $B(x)$ with respect to $x$ explicitly as
\begin{equation}
\frac{dB(x)}{dx} = 2\sum_{k=0}^{t-i}\frac{1}{k!}\sum_{l=0}^{\tau(i,j)+k}{\tau(i,j)+k\choose l}a^l\frac{d\left(x^k(\sqrt{x})^{\theta(i,j)+l-k+1}K_{\theta(i,j)+l-k+1}(2\sqrt{x})\right)}{dx}.
\end{equation}
With the help of the differential property of Bessel-K function
\begin{equation}
\frac{d(x^vK_v(x))}{dx} = -x^vK_{v-1}(x),
\end{equation}
we get
\begin{multline}
\frac{d\left(x^k(\sqrt{x})^{\theta(i,j)+l-k+1}K_{\theta(i,j)+l-k+1}(2\sqrt{x})\right)}{dx}\\
=kx^{k-1}x^{\frac{\theta(i,j)+l-k+1}{2}}K_{\theta(i,j)+l-k+1}(2\sqrt{x})-x^kx^{\frac{\theta(i,j)+l-k}{2}}K_{\theta(i,j)+l-k}(2\sqrt{x}).
\end{multline}
As a result, some simple algebraic manipulations yield the desired result.

\section{Proof of Theorem \ref{theorem:asy2}}\label{appendix:theorem:asy2}
When $n_s = 1$, $\lambda_{\sf max}$ is statistically equivalent to ${\bf h}^{\dag}{\bf \Delta}{\bf h}$, where ${\bf h}\in {\cal CN}^{q\times 1}$ and ${\bf \Delta}= {\sf diag}\{\beta_1,\dots,\beta_q\}$ is a $q\times q$ diagonal matrix with $\beta_i$, for $i=1,\dots,q$ being the $q$ non-zero eigenvalues of the matrix $\mathbf{H}_2^{\dag}\left(a\mathbf{H}_2\mathbf{H}_2^{\dag}+\mathbf{I}_{n_d}\right)^{-1}\mathbf{H}_2$.

Conditioned on ${\bf \Delta}$, $\lambda_{\sf max}$ is the sum of $q$ independent chi-square random variables. Hence, the near zero behavior of $\lambda_{\sf max}$ can be expressed as \cite{A.Ribeiro}
\begin{equation}
f_{\lambda_{\sf max}|{\bf \Delta}}(x)\approx \frac{1}{\Gamma(q)}\prod_{i=1}^{q}\beta_i^{-1}x^{q-1}.
\end{equation}
For this reason, the unconditional asymptotic expansion can be obtained by averaging over ${\bf \Delta}$. Applying the joint pdf of $\beta_i$ presented in \cite{S.Jin}, we have
\begin{multline}
f_{\lambda_{\sf max}}(x)\approx \\
\frac{x^{q-1}}{\Gamma(q)\prod_{i=1}^{q}\Gamma(p-i+1)\Gamma(q-i+1)}\int_0^\frac{1}{a}\cdots\int_0^\frac{1}{a}\prod_{i<j}^q(\beta_j-\beta_i)^2 \prod_{i=1}^{q}\frac{\beta_i^{p-q-1}e^{-\frac{\beta_i}{1-a\beta_i}}}{(1-a\beta_i)^{p+q}}d\beta_1\cdots\beta_q,
\end{multline}
where the multiple integrals can be solved using the technique in \cite{M.Chiani}. Hence,
\begin{equation}
f_{\lambda_{\sf max}}(x)\approx \frac{x^{q-1}\det({\bf X})}{\Gamma(q)\prod_{i=1}^{q}\Gamma(p-i+1)\Gamma(q-i+1)},
\end{equation}
where ${\bf X}$ is a $q\times q$ matrix with entries
\begin{equation}
[{\bf X}]_{i,j} = \int_0^\frac{1}{a}\beta^{p-q+i+j-3}e^{-\frac{\beta}{1-a\beta}}(1-a\beta)^{-p-q}d\beta.
\end{equation}
Explicitly evaluating the above integral yields the desired result.

\section{Proof of Theorem \ref{theorem:asy1}}\label{appendix:theorem:asy}
To obtain the asymptotic first order expansion, we find it more convenient to start from the following alternative equivalent pdf expression of $\lambda_{\sf max}$ presented in \cite{S.Jin}.
\begin{equation}\label{eqn:pdff11}
f_{\lambda_{\sf max}}(x) = \frac{2x^{n_s-1}e^{-ax}}{\Gamma(p)\Gamma(n_s)}\sum_{i=0}^{n_s}{n_s\choose i}a^ix^{\frac{p-n_s+i}{2}}K_{p-n_s+i}(2\sqrt{x}),
\end{equation}
which can be written as a Maclaurin series, and we are interested in the first term with the minimum exponent and non-zero coefficient. To achieve this, we consider three separate cases:

1) $p>n_s$. We first notice that the minimum exponent for $x^{n_s-1}e^{-ax}$ is $n_s-1$. Hence, the next key step is to determine the minimum exponent of the summation in (\ref{eqn:pdff11}). When $p>n_s$, it is easy to see that $p-n_s+i>0$,  for $i = 0, \dots, n_s$. Then, based on the asymptotic series expansion of $K_v(x)$, we have
\begin{multline}\label{eqn:asympt1}
x^{\frac{p-n_s+i}{2}}K_{p-n_s+i}(2\sqrt{x})= \frac{1}{2}\sum_{k=0}^{p-n_s+i-1}\frac{\Gamma(p-n_s+i-k)}{\Gamma(k+1)}\left(-x\right)^k\\
-\frac{(-x)^{p-n_s+i}}{2}\sum_{k=0}^{\infty}\frac{\ln x-\psi(k+1)-\psi(p-n_s+i+k+1)}{\Gamma(k+1)\Gamma(p-n_s+i+k+1)}x^k.
\end{multline}
Clearly, the dominant term is given by $k = 0$ in the first term, i.e.,
\begin{equation}
x^{\frac{p-n_s+i}{2}}K_{p-n_s+i}(2\sqrt{x}) \approx \frac{\Gamma(p-n_s+i)}{2}.
\end{equation}
Therefore,
\begin{equation}
f_{\lambda_{\sf max}}(x) \approx  \frac{x^{n_s-1}}{\Gamma(p)\Gamma(n_s)}\sum_{i=0}^{n_s}{n_s\choose i}a^i \Gamma(p-n_s+i).
\end{equation}

2) $p = n_s$. In this case, the summation in (\ref{eqn:pdff11}) can be expressed as
\begin{equation}
\sum_{i=0}^{n_s}{n_s\choose i}a^ix^{\frac{i}{2}}K_{+i}(2\sqrt{x})= K_0(2\sqrt{x})+\sum_{i=1}^{n_s}{n_s\choose i}a^ix^{\frac{i}{2}}K_{i}(2\sqrt{x}).
\end{equation}
The dominant term can be found using (\ref{eqn:asympt1}) and the following asymptotical expansion of $K_0(x)$
\begin{equation}
K_0(x) \approx  -c -\ln\frac{x}{2},
\end{equation}
where $c$ is the Euler constant, and we therefore have
\begin{equation}
f_{\lambda_{\sf max}}(x) \approx \frac{x^{n_s-1}}{\Gamma(p)\Gamma(n_s)}\left(-c-\ln x +\sum_{i=1}^{n_s}{n_s\choose i}a^i \Gamma(p-n_s+i)\right).
\end{equation}

3) $p < n_s$. Since the modified Bessel function of second kind is a symmetric function with respect to $v$, i.e., $K_v(x) = K_{-v}(x)$, we can express the summation in (\ref{eqn:pdff11}) as
\begin{multline}
\sum_{i=0}^{n_s}{n_s\choose i}a^ix^{\frac{p-n_s+i}{2}}K_{p-n_s+i}(2\sqrt{x})
 =\sum_{i=0}^{n_s-p-1}{n_s\choose i}a^ix^{\frac{p-n_s+i}{2}}K_{n_s-p-i}(2\sqrt{x})+\\
 {n_s\choose n_s-p}a^{n_s-p}K_0(2\sqrt{x})+\sum_{i=n_s-p+1}^{n_s}{n_s\choose i}a^ix^{\frac{p-n_s+i}{2}}K_{p-n_s+i}(2\sqrt{x}).
\end{multline}
A close observation shows that the term with minimum exponent comes from the first part in the right hand side of the above equation. Hence, as $x\rightarrow 0$, we have
\begin{align}
\sum_{i=0}^{n_s}{n_s\choose i}a^ix^{\frac{p-n_s+i}{2}}K_{p-n_s+i}(2\sqrt{x})
&\approx \sum_{i=0}^{n_s-p-1}x^{n_s-p+i}{n_s\choose i}a^ix^{\frac{n_s-p-i}{2}}K_{n_s-p-i}(2\sqrt{x})\nonumber\\
&\approx \sum_{i=0}^{n_s-p-1}x^{n_s-p+i}{n_s\choose i}a^i\frac{\Gamma(n_s-p-i)}{2}.
\end{align}
Clearly, the dominant term with minimum exponent is associated with $i=0$. Hence, we have
\begin{equation}
\sum_{i=0}^{n_s}{n_s\choose i}a^ix^{\frac{p-n_s+i}{2}}K_{p-n_s+i}(2\sqrt{x})\approx \frac{\Gamma(n_s-p)}{2}x^{p-n_s}.
\end{equation}
Therefore, we get
\begin{equation}
f_{\lambda_{\sf max}}(x) \approx \frac{\Gamma(n_s-p)}{\Gamma(p)\Gamma(n_s)}x^{p-1}.
\end{equation}

\section{Proof of Theorem \ref{theorem:asyantenna1}}\label{appendix:theorem:asyantenna1}
When $n_s\rightarrow\infty$, by law of large numbers, we have
\begin{equation}
\lim_{n_s\rightarrow\infty}\frac{{\bf H}_1{\bf H}_1^{\dag}}{n_s} = {\bf I}.
\end{equation}
Noting that $\lambda_{\sf max}$ is the maximum eigenvalue of $\mathbf{H}_2^{\dag}\left(a\mathbf{H}_2\mathbf{H}_2^{\dag}+\mathbf{I}_{n_d}\right)^{-1}\mathbf{H}_2\mathbf{H}_1\mathbf{H}_1^{\dag}$, $\lambda_{\sf max}$ is equivalent to the maximum eigenvalue of $n_s\mathbf{H}_2^{\dag}\left(a\mathbf{H}_2\mathbf{H}_2^{\dag}+\mathbf{I}_{n_d}\right)^{-1}\mathbf{H}_2$, which completes the first claim.

When $n_d\rightarrow\infty$, by law of large numbers, we have
\begin{equation}
\lim_{n_d\rightarrow\infty}\frac{{\bf H}_2^{\dag}{\bf H}_2}{n_d} = {\bf I}.
\end{equation}
Then, we see that $\mathbf{H}_2^{\dag}\left(a\mathbf{H}_2\mathbf{H}_2^{\dag}+\mathbf{I}_{n_d}\right)^{-1}\mathbf{H}_2 \approx \frac{n_d}{an_d+1}{\bf I}$, and the second claim follows immediately.

When $n_r\rightarrow\infty$, by law of large numbers, we have
\begin{equation}
\lim_{n_r\rightarrow\infty}\frac{{\bf H}_2{\bf H}_2^{\dag}}{n_r} = {\bf I},
\end{equation}
and a close observation reveals that $\mathbf{H}_2^{\dag}\left(a\mathbf{H}_2\mathbf{H}_2^{\dag}+\mathbf{I}_{n_d}\right)^{-1}\mathbf{H}_2$ has only $n_d$ non-zero eigenvalues which is identical to $\frac{n_r}{an_r+1}$. Thus, applying \cite[Lemma 1]{C.Zhong} yields the desired result.

\section{Proof of (\ref{eqn:poweroffset})}\label{appendix:eqn:poweroffset}
From the definition, we can compute the high SNR power offset as
\begin{equation}
{\cal L}_{\infty} = \log_2\frac{n_r}{k}-{\tt E}\{\log_2\lambda_{\sf max}\}.
\end{equation}
When $q=1$, the maximum eigenvalue can be alternatively expressed as
\begin{equation}
\lambda_{\sf max}=\frac{{\bf h}_1^{\dag}{\bf h}_1{\bf h}_2^{\dag}{\bf h}_2}{a{\bf h}_2^{\dag}{\bf h}_2+1},
\end{equation}
where ${\bf h}_1\in {\cal CN}^{n_s\times 1}$ and ${\bf h}_2\in {\cal CN}^{p\times 1}$. Hence, we have
\begin{equation}
{\tt E}\{\log_2\lambda_{\sf max}\}={\tt E}\{\log_2({\bf h}_1^{\dag}{\bf h}_1)\}+{\tt E}\{\log_2({\bf h}_2^{\dag}{\bf h}_2)\}-{\tt E}\{\log_2(a{\bf h}_2^{\dag}{\bf h}_2+1)\}.
\end{equation}
As such, the desired result can be obtained by using the results presented in \cite{A.Grant,H.Shin}.

\nocite{*}
\bibliographystyle{IEEE}
\begin{footnotesize}

\end{footnotesize}

\begin{figure}[htb!]
\centering
\includegraphics[scale=1]{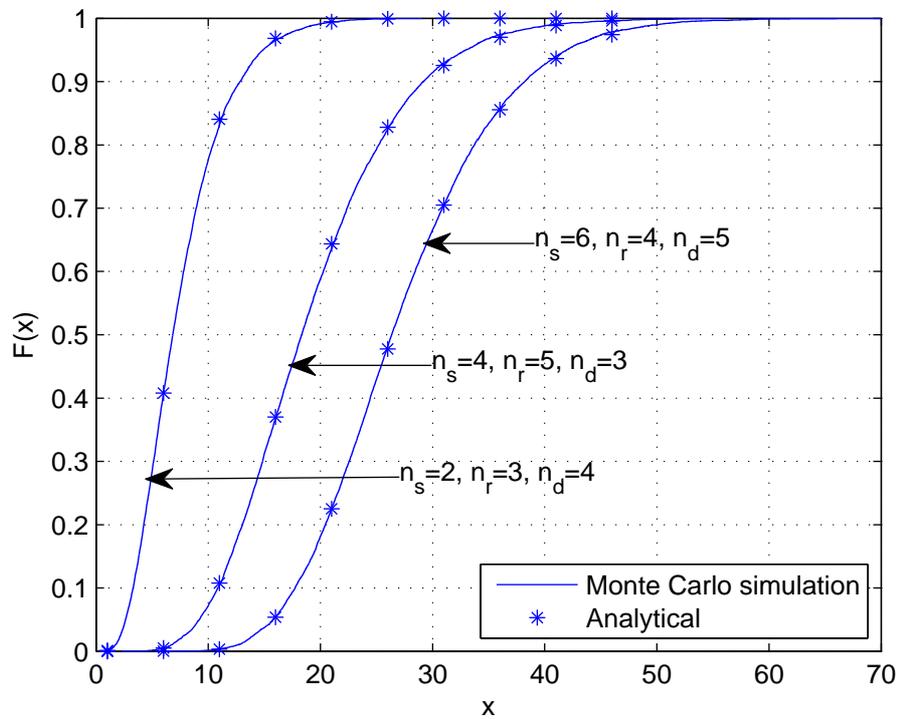}
\caption{The cdf results of $\lambda_{\sf max}$: Analytical versus Monte Carlo simulations.}\label{fig:fig1}
\end{figure}

\begin{figure}[htb!]
\centering
\includegraphics[scale=0.85]{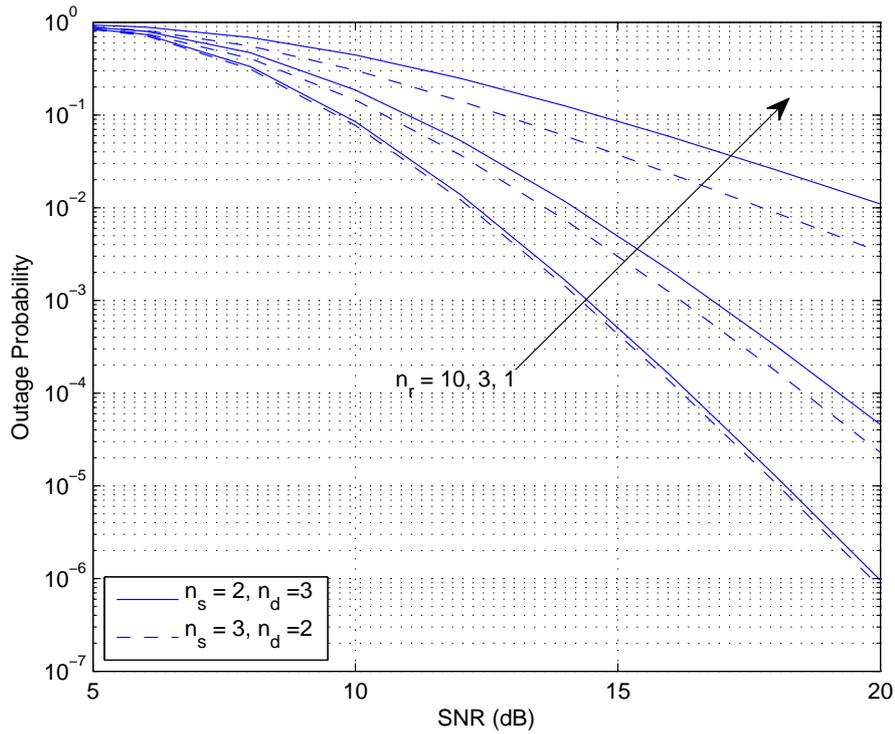}
\caption{Impact of $n_r$ on the outage probability of optimal beamforming MIMO dual-hop AF systems when $\alpha = \rho$.}\label{fig:fig21}
\end{figure}

\begin{figure}[htb!]
\centering
\includegraphics[scale=0.85]{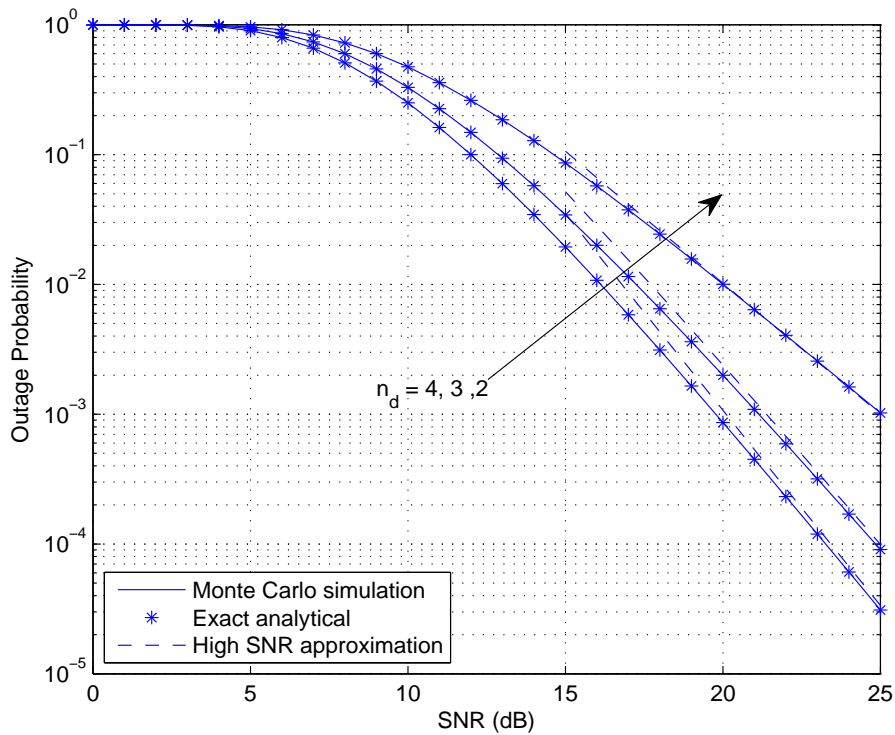}
\caption{Outage probability results of optimal beamforming MIMO dual-hop AF systems when $n_s = 3$, $n_r = 1$, and $\alpha = 0.5\rho$.}\label{fig:fig2}
\end{figure}

\begin{figure}[htb!]
\centering
\includegraphics[scale=0.85]{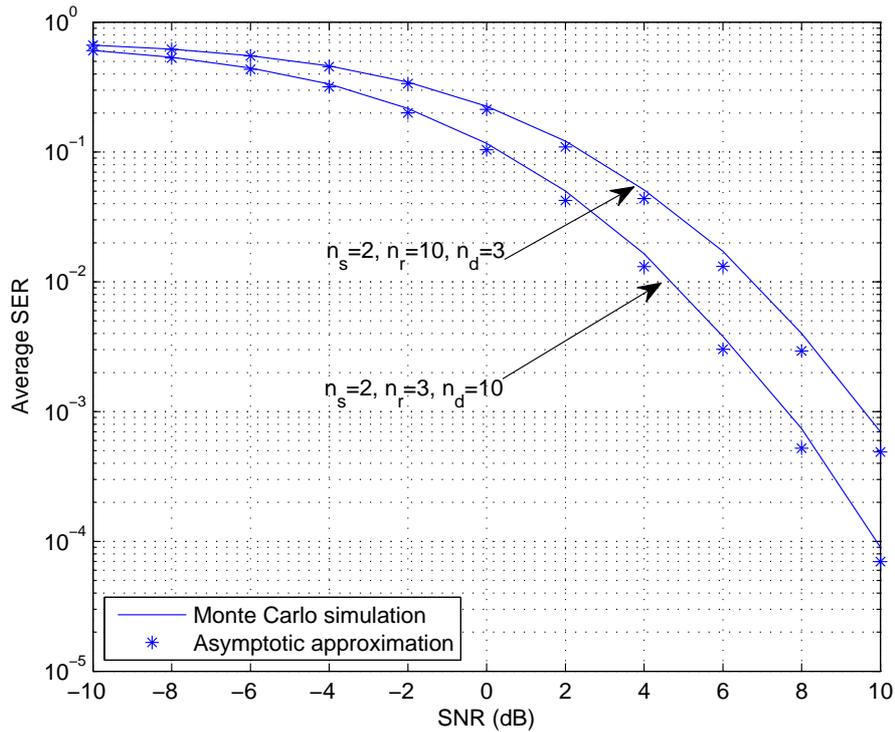}
\caption{Average SER of optimal beamforming MIMO dual-hop systems when $\alpha = \rho$: Exact versus asymptotic approximations.}\label{fig:fig3}
\end{figure}

\begin{figure}[htb!]
\centering
\includegraphics[scale=0.85]{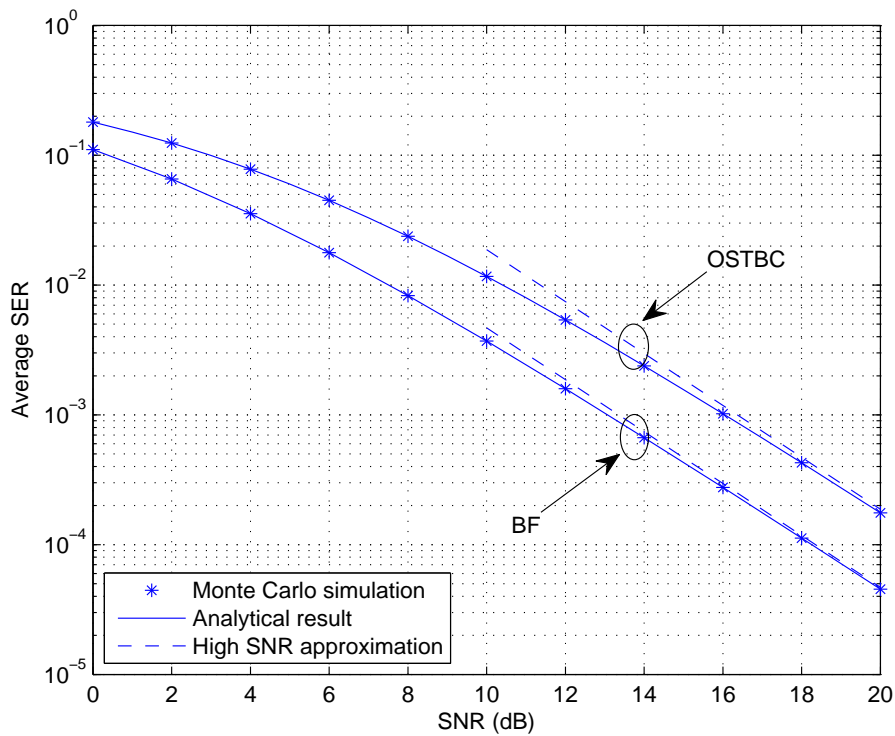}
\caption{Average SER of optimal beamforming MIMO dual-hop systems when $n_s =2$, $n_r =1$, $n_d =3$, and $\alpha = \rho$, with BPSK modulation: Beamforming versus OSTBC.}\label{fig:fig4}
\end{figure}

\begin{figure}[htb!]
\centering
\includegraphics[scale=0.85]{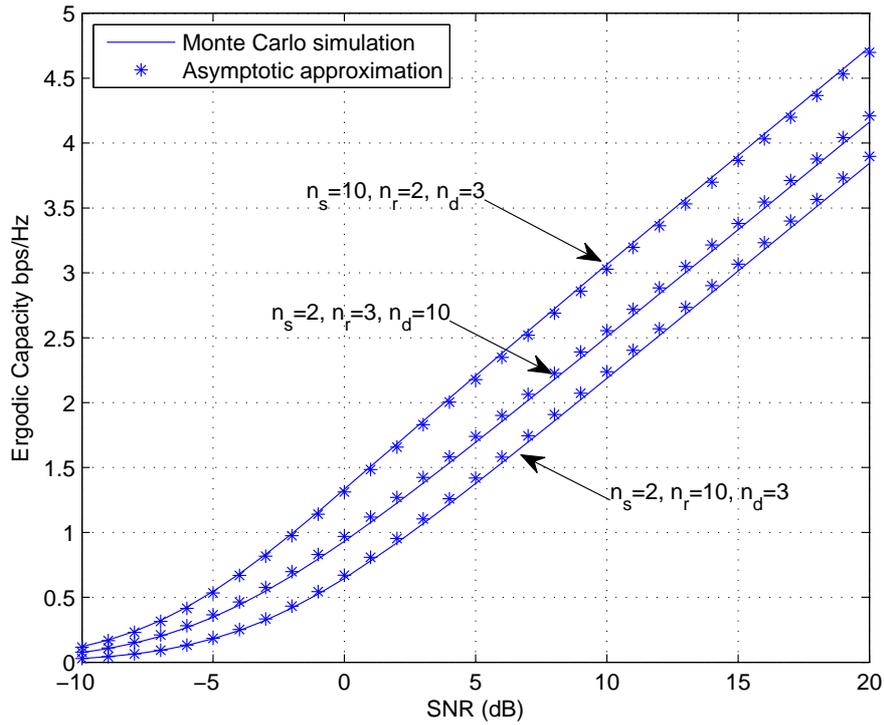}
\caption{Ergodic capacity of optimal beamforming MIMO dual-hop AF systems when $\alpha = \rho$: Exact versus asymptotic approximations.}\label{fig:fig5}
\end{figure}

\begin{figure}[htb!]
\centering
\includegraphics[scale=1]{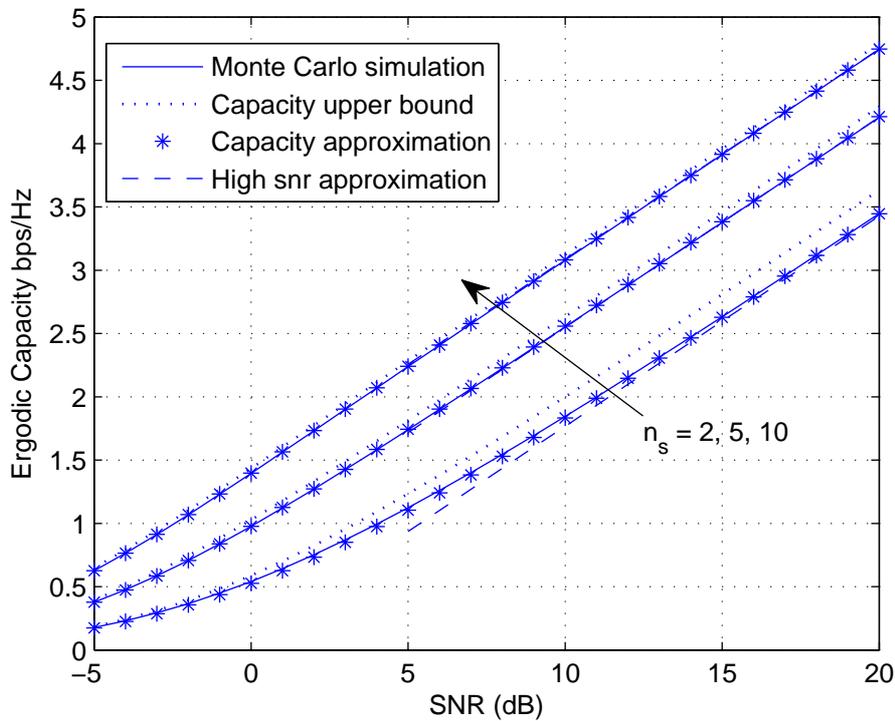}
\caption{Ergodic capacity of optimal beamforming  MIMO dual-hop AF systems for different $n_s$, when $n_r=1$, $n_d=4$, and $\alpha = \rho$.}\label{fig:fig6}
\end{figure}

\end{document}